# Toward Coupled-Cluster Implementations in Nuclear Structure

D.J. Dean[1] and M. Hjorth-Jensen[2]

[1]*Physics Division, Oak Ridge National Laboratory*
*P.O. Box 2008, Oak Ridge, TN 37831-6373 USA*
[2]*Department of Physics, University of Oslo, N-0316 Oslo, Norway*

**Abstract.** We discuss an initial implementation of the coupled-cluster method for nuclear structure calculations and apply our method to $^4$He. We will discuss the future directions that this research will take as we move from method testing and development to more complete calculations.

## THE NUCLEAR MANY-BODY PROBLEM AND COUPLED CLUSTER THEORY

Within the next 15 years the Rare Isotope Accelerator (RIA) will become a reality in some form. Even today, we are facing, and will continue to face, an explosion of nuclear structure data that will require a confrontation with nuclear theory both to verify the theory and to guide future experiments at RIA and beyond. In spite of the vast number of nuclei that will be reached with RIA and other next-generation rare-isotope accelerators, many nuclei will remain experimentally unexplored, and reliable theoretical models will need to be developed to predict their properties.

One may follow various theoretical paths to obtain information about the properties of nuclear systems. One path originates from following a reductionist approach. One begins with some derivation of the nucleon-nucleon interaction (such as that built upon meson exchange, chiral perturbation theory, or phenomenology), and one develops computational tools for solving the many-body problem, as well as one can from this point of view. Examples of research efforts pursuing this approach include the Green's Function Monte Carlo collaboration who begin with the Argonne interaction and supplement it with effective three-body interactions [1], and the no-core shell-model collaboration who generate a G-matrix+folded-diagrams effective interaction and diagonalize in a given model space [2]. Both methods are *ab initio* from the many-body point of view: they begin with the bare nucleon-nucleon interactions. Another valid approach which has been successful requires the development of effective nuclear interactions at either the mean-field level through the use of Skyrme-like forces (see, for example, Ref. [3] and references therein), or with the shell model by using effective interactions derived from experimental level and transition information [4].

The *ab initio* approaches, while difficult, allow one to study emergent phenomena such as deformation or vibrations of the nuclear systems from the fundamental level of the bare interactions; however, the applications of these methods are at the present time limited to light nuclear systems. The effective interactions (whether of the mean-field or shell-model variety) may be applied to various regions of the nuclear chart, but they often (especially in shell-model applications) rely on data-fitting within the region being calculated. The successful shell-model interactions, such as the $1s0d$ interaction [5] or the $0f1p$ interactions [6], all require large data sets in order to be adjusted appropriately to reproduce existant data and predict certain quantities within a given region. Herein lies the difficulty of relying on fitted interactions: the experimental data coming from RIA and other facilities will not be dense enough to allow for a successful fitting of effective shell-model interactions in regions of interest. With this in mind, it becomes essential for nuclear theorists to explore methods that will allow for *ab initio* calculations of nuclei both near stability and in regions where RIA and other radio-isotopic facilities will probe.

In these Proceedings, we will discuss a many-body approach, known as coupled-cluster theory, that may prove quite useful in applications to nuclear structure. Coupled-cluster theory was first introduced in nuclear physics by Coester [7] and Coester and Kummel [8]. Initial nuclear structure applications came in the mid-1970s with several papers from the Bochum group [9]. Since that time, nuclear physics applications have been rather sporadic. On the other hand, the history of coupled-cluster implementations in computational quantum chemistry have been rather dramatic. The first chemistry application was discussed by Čížek and Paldus [10], and the method became computationally feasible due to work by Pople [11] and Bartlett and Purvis [12]. Pople received the Nobel Prize in chemistry in 1998 for his contributions to computational quantum chemistry algorithms such as those developed for the coupled-cluster approach. (See Ref. [13] for an excellent practitioner's review on applications in quantum chemistry.) The interesting and desirable theoretical properties of the coupled-cluster method within computational chemistry have made it *the* method of choice in computations of many-body correlation effects in atomic, molecular, and chemical systems. While it was originally developed for the many-body ground-state, applications of the coupled-cluster method in quantum chemistry now extend to excited states and open-shell systems.

Recent nuclear applications of the coupled-cluster technique include approaches in coordinate-space being addressed by the Manchester group [14]. Recently, Heisenberg and Mihaila [15] have suggested a somewhat different formulation than that espoused in quantum chemistry. Outside of these efforts, only sporadic development work within coupled-cluster methods has been pursued within nuclear theory. The reasons for this lack of pursuit in the nuclear many-body problem probably arise from the lack of a good starting point, that is, a good bare nucleon-nucleon interaction. In the last 10 years this problem has been effectively eliminated due to excellent nucleon-nucleon interactions that give $\chi^2$ per degree-of-freedom of nearly one. These interactions include the phenomenological Argonne $V_{18}$ potential [16], the meson exchange potentials such as CD-Bonn[17], and the very nucleon-nucleon potentials based on chiral perturbation theory [18]. Another reason that the method was not pursued was certainly the lack of computational power available in the late 1970s as compared to today.

Several salient features of coupled-cluster theory make it an attractive theory to pursue. It is a fully microscopic theory that can be used to obtain energies and eigenstates of a given Hamiltonian. Furthermore, the theory is capable of systematic improvements through increasingly higher-order implementations of a well-defined scheme of hierarchical approximations. Coupled-cluster theory is size extensive, which means that only linked diagrams enter into a given computation. This is not true in typical shell-model particle-hole truncation schemes [13]. The method is also size consistent: the energy of two non-interacting fragments computed separately is the same as that computed for both fragments simultaneously [19]. This property is particularly relevant for chemical reaction studies and is not a property of the shell model.

## APPROACH TO COUPLED-CLUSTER THEORY

Our approach to the coupled-cluster equations contains two steps. We first derive an effective nucleon-nucleon potential through a *G*-matrix formulation. We then apply the coupled-cluster method to the Hamiltonian containing this *G*-matrix.

### The *G*-matrix

The presence of a hard core in various channels of the nucleon-nucleon interaction (with repulsion on the order of $2-5$ GeV) causes difficulty for theories that wish to use basis state expansion techniques. One way to overcome this difficulty is to use a renormalized effective interaction within the model space where one will actually perform computations. This model-space, dubbed the *P* space, is a subset of the full Hilbert space. The excluded space, dubbed the *Q* space, represents the remaining part of the Hilbert space and is, in principle, infinite in size. One may project any operator into the *P* or *Q* space through the use of projection operators, $\hat{P}$ and $\hat{Q}$, such that $\hat{P}+\hat{Q}=1$. Brueckner [20] originally developed the *G*-matrix theory that allows for the solution of the full *A*-body problem in the reduced Hilbert space. See [21], and references therein for full details. Here we only discuss the basic equation which is

$$G(\omega) = V + V\frac{Q}{\omega - QTQ}G(\omega),\qquad(1)$$

where *V* is the bare nucleon-nucleon interaction, *T* is the kinetic energy operator, and $\omega$ is a starting energy. We discuss the determination of the starting energy below. This equation requires iteration. Diagrammatically, this amounts to generating all particle-hole ladder diagrams, with intermediate two-particle states outside the *P*-space, to infinite order.

We demonstrate how our calculation proceeds in Fig. 2. We first choose the *P* space, as shown in Fig. 2. Within that space, we compute the *G*-matrix elements of the renormalized interactions. We then define a reference Slater determinant from which we perform the coupled-cluster calculation. By performing the coupled-cluster calculations only in the $P$-space, we insure that no double counting of many-body perturbation theory diagrams occurs. Those diagrams that we do not include in this expansion are those for which a particle below the Fermi energy in the reference Slater determinant moves to the $Q$-space; however, as one increases the $P$-space, the contribution of these diagrams to observable quantities such as the energy should become very small.

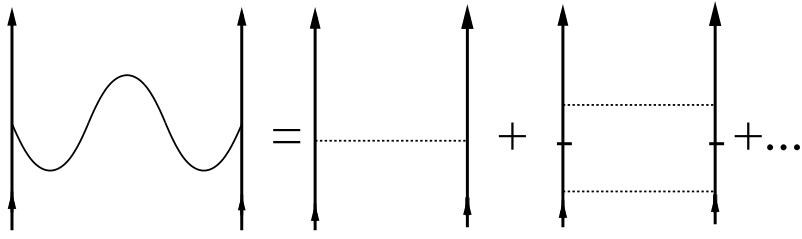

**FIGURE 1.** The diagrams summed by the $G$-matrix renormalization of the bare nucleon-nucleon potential. The wavy line represents the medium-dependent reaction matrix $G$. Physically, these diagrams mean that the particles must interact virtually with each other an arbitrary number of times in order to produce a finite interaction matrix element. Railed lines denote fermions with momentum greater than $k_f$ (residing in the $Q$ space).

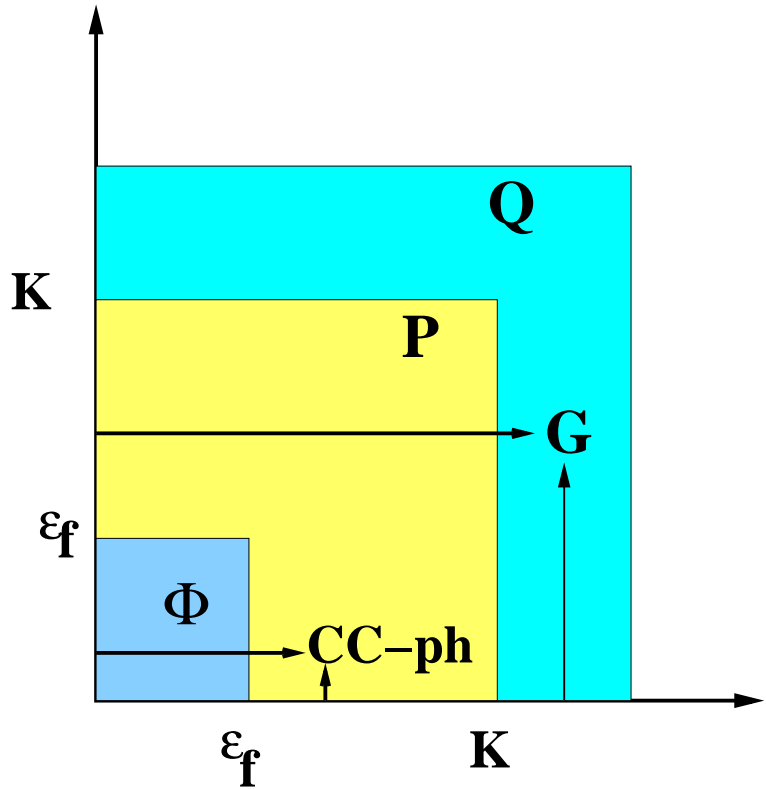

**FIGURE 2.** The choice of model space. Particle-hole excitations from the $P$-space (with energy cutoff $K$) to the $Q$-space are allowed during the computation of the $G$-matrix. Coupled-cluster computations occur only in the $P$-space where the Fermi energy, $\varepsilon_f$, is determined by the reference Slater determinant $|\Phi\rangle$.

By implementing the $G$-matrix formalism, we obtain as our Hamiltonian

$$H = \sum_{pq} K_{pq} a_p^\dagger a_q + \frac{1}{4} \sum_{pqrs} \langle pq | G | rs \rangle a_p^\dagger a_q^\dagger a_s a_r , \qquad (2)$$

where $K_{pq}$ are the one-body matrix elements of the kinetic energy operator, $K_{pq} = \langle \phi_p | K | \phi_q \rangle$, and $\langle pq | G(\omega) | rs \rangle$ are the antisymmetrized two-body matrix elements of the effective nucleon-nucleon interaction. The single-particle wave functions are the basis states of the problem, and the labels $p, q, r, s$ represent all single-particle quantum numbers. In the oscillator basis, $p = \{n, l, j, m, t_z\}$, where $n$ is the principal quantum number, $l$ is the particle angular momentum, $j$ is the total angular momentum, $m$ is the angular momentum projection, and $t_z$ is the isospin projection of the particle. In the following, we use the labels $ijk$ to represent single-particle states below the Fermi surface, and labels $abc$ to indicate single-particle states above the Fermi surface.

The Hamiltonian may be written in a slightly more convenient form by explicitly calculating the expectation of the Hamiltonian in the reference state $| \Phi \rangle$, $E_0 = \langle \Phi | H | \Phi \rangle$. This reference state is a single Slater determinant and represents, in this work, a doubly closed shell system. In this case, the Hamiltonian becomes

$$H = \sum_{pq} f_{pq} \{a_p^\dagger a_q\} + \frac{1}{4} \langle pq | G | rs \rangle \{a_p^\dagger a_q^\dagger a_s a_r\} + E_0 , \tag{3}$$

where the $\{\}$ indicates normal ordering relative to the Fermi vacuum. The Fock operator is given by

$$f_{pq} = \langle p | K | q \rangle + \sum_i \langle pi | G | qi \rangle . \tag{4}$$

## Our implementation of the coupled-cluster method

The basic idea of coupled-cluster theory is that the correlated many-body wave function $| \Psi \rangle$ may be obtained by application of a correlation operator, $T$, such that

$$| \Psi \rangle = \exp(-T) | \Phi \rangle , \tag{5}$$

where $\Phi$ is a reference Slater determinant chosen as a convenient starting point. For example, we use the filled $0s$ state as the reference determinant for $^4$He. This exponential ansatz has been well justified for many-body problems using a formalism in which the cluster functions are constructed by cluster operators acting on a reference determinant [22].

The correlation operator $T$ is given by

$$T = T_1 + T_2 + \cdots T_A , \tag{6}$$

and represent various $n$-particle-$n$-hole ($n$p-$n$h) excitation amplitudes such as

$$T_1 = \sum_{a<\varepsilon_f, i>\varepsilon_f} t_i^a a_a^\dagger a_i , \tag{7}$$

$$T_2 = \frac{1}{4} \sum_{i,j<\varepsilon_f; ab>\varepsilon_f} t_{ij}^{ab} a_a^\dagger a_b^\dagger a_j a_i , \tag{8}$$

and higher-order terms for $T_3$ to $T_A$. We are currently exploring the coupled-cluster method at the $T_1$ and $T_2$ level. This is commonly referred to in the literature as Coupled-Cluster Singles and Doubles (CCSD).

We compute the expectation of the energy from

$$E = \langle \Psi_0 | \exp(-T) H \exp(T) | \Psi_0 \rangle . \tag{9}$$

The Baker-Hausdorf relation may be used to rewrite the similarity transformation as

$$\exp(-T) H \exp(T) = H + [H, T_1] + [H, T_2] + \frac{1}{2} [[H, T_1], T_1] + \frac{1}{2} [[H, T_2], T_2] + [[H, T_1], T_2] + \cdots . \tag{10}$$

The expansion terminates exactly at four nested commutators when the Hamiltonian contains, at most, two-body terms, and at six-nested commutators when three-body potentials are present. We stress that this termination is exact, thus allowing for a derivation of exact expressions for the $T_1$ ($1p$-$1h$) and $T_2$ ($2p$-$2h$) amplitudes. The notation $t_n$ will denote all of the amplitudes of a given class, so that $t_1$ represents the set of all $t_i^a$ amplitudes, and a similar definition holds for $t_2$. The equations for amplitudes are found by left projection of excited Slater determinants so that

$$0 = \langle \Phi_i^a | \exp(-T) H \exp(T) | \Phi \rangle , \tag{11}$$

$$0 = \langle \Phi_{ij}^{ab} | \exp(-T) H \exp(T) | \Phi \rangle . \tag{12}$$

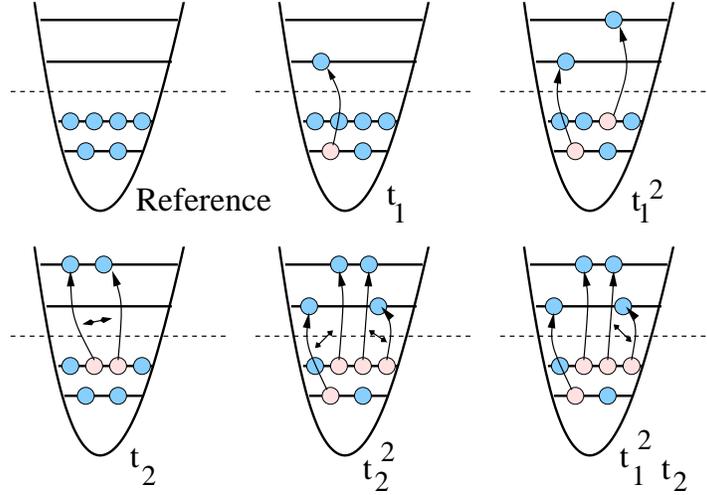

**FIGURE 3.** Shown are the various types of correlations available in the CCSD equations. Upper left: the reference Slater determinant. Upper middle: $1p$-$1h$ excitations. Upper right: two independent $1p$-$1h$ excitations. CCSD allows up to four independent $1p$-$1h$ excitations. Lower right: a single correlated $2p$-$2h$ excitation. Lower middle: two $2p$-$2h$ excitations. Lower right: two $1p$-$1h$ excitations and a $2p$-$2h$ excitation.

The commutators also generate nonlinear terms within these expressions. To derive these equations is straightforward, but tedious, work [13]. While the resulting equations for the single and double excitation amplitudes appear quite lengthy, they are solvable through iterative techniques. We show in Fig. 3 an illustration of the types of excitations allowed from the CCSD equations.

Once the amplitudes are obtained, the energy of the system may then be calculated. The CCSD energy is

$$\langle H \rangle = E_{CCSD} = \sum_{ia} f_{ia} t_i^a + \frac{1}{4} \sum_{aibj} \langle ij | G | ab \rangle t_{ij}^{ab} + \frac{1}{2} \sum_{aibj} \langle ij | G | ab \rangle t_i^a t_j^b + E_0 \,. \qquad (13)$$

This equation is not restricted to the CCSD approximation. Since higher-order excitation operators such as $T_3$ and $T_4$ cannot produce fully contracted terms with the two-body Hamiltonian, their contribution to the energy equation is zero. Higher-order excitation clusters can contribute indirectly to the energy through the equations used to determine the amplitudes.

Because the energy is computed using projective, asymmetric techniques, an important question concerns the physical reality of the coupled-cluster energy. Quantum mechanics requires that physical observables should be expectation values of Hermitian operators. The coupled-cluster energy expression contains the non-Hermitian operator $\exp(-T)H\exp(T)$. However, if $T$ is not truncated, the similarity-transformed operator exhibits an energy-eigenvalue spectrum that is identical to the original Hermitian operator, H, thus justifying its formal use in quantum-mechanical models. From a practical point of view, the coupled-cluster energy tends to follow the expectation value result (if the theory is reformulated as a variational theory), even when $T$ is truncated.

We note that the nonlinear terms that appear within the amplitude equations include terms that allow for $4p$-$4h$ excitations. Indeed, while we speak of doubles in terms of amplitudes, the class of many-body diagrams involved in the theory includes fourth-order diagrams.

Because of the nonlinearity of the equations, one must have a good first guess for the $np$-$nh$ amplitudes. To solve the equations, we rewrite them in the following form

$$D_{ai} t_i^a = f_{ai} + F_1(t_1, t_2) \,, \qquad (14)$$

$$D_{abij} t_{ij}^{ab} = \langle ij | G | ab \rangle + F_2(t_1, t_2) \,, \qquad (15)$$

where $D_{ia} = f_{ii} - f_{aa}$, and $D_{ijab} = f_{ii} + f_{jj} - f_{aa} - f_{bb}$ are energy denominators. We have made a simple rearrangement of the projection equations for the amplitudes, Eqs. 12, in terms of these denominators. The functions $F_1$ and $F_2$ represent all other terms in the amplitude equations and depend on the amplitudes $t_1$ and $t_2$. For closed-shell nuclei, we use a Moller-Plesset-like approach to generate the first guess for the iteration: we assume that all initial amplitudes

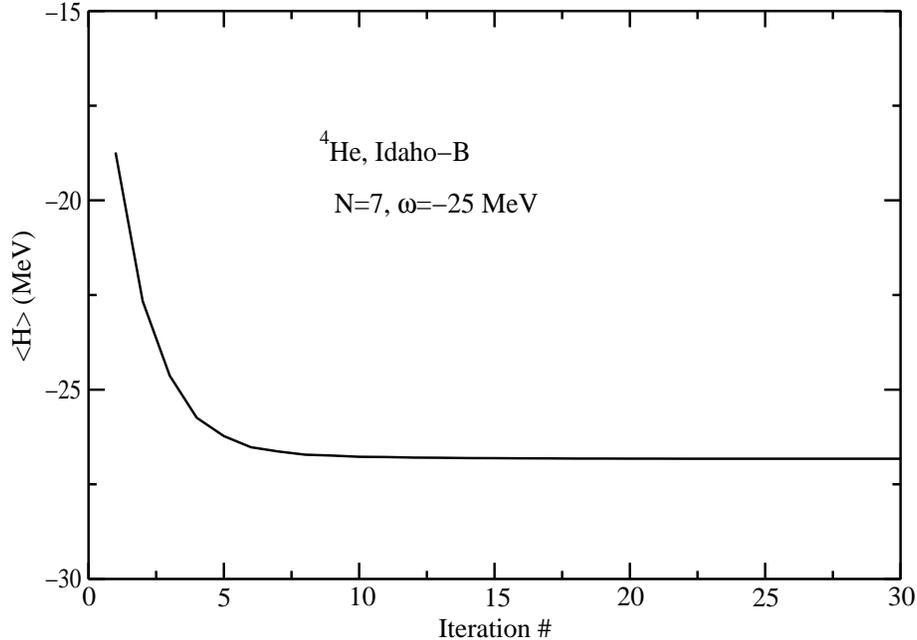

**FIGURE 4.** The convergence of the ground-state energy as a function of the CCSD iterations.

on the right-hand side of Eqs. 14 and 15 are zero. This yields

$$t_i^a(1) = \frac{f_{ai}}{D_{ai}}, \tag{16}$$

$$t_{ij}^{ab}(1) = \frac{<ab\,|\,G\,|\,ij>}{D_{ijab}}. \tag{17}$$

We use this first guess for $t_1(1)$ and $t_2(1)$ to calculate the right-hand side of Eqs. 14 and 15. This then yields a new set of amplitudes $t_1(2)$ and $t_2(2)$ on the left-hand side of Eqs. 14 and 15. We are then able to iterate the solution to convergence.

Shown in Fig. 4 is the convergence of the total energy of the system as a function of iteration number. For our test example, $^4$He, we achieve convergence at the $10^{-5}$ level by 30 iterations in a model-space that includes seven major oscillator shells. Notice from the figure that most of the convergence is obtained within 10 iterations.

By investigating the different terms within the equations and their contributions to the energies, one is able to generate a correspondence between CCSD and many-body perturbation theory. One finds that CCSD iterates the lowest first-, second-, third-, and fourth-order many-body perturbation theory diagrams to all orders. It should be noted that the third-order diagrams are incomplete at the CCSD level of truncation, although third-order corrections may be included if they are desired [23].

## INITIAL RESULTS

Our overall goals are to understand the structure of nuclei using coupled-cluster theory as our tool. We are at the very beginning of this effort and have a few preliminary results that we will report here. We are computing at the singles and doubles level of the coupled-cluster theory. At this level of truncation, we assume that all $t_3$ and higher-order amplitudes are zero. We also assume for the moment that only two-body potentials are present in the nuclear problem. We have not yet corrected these results for center-of-mass contamination, which means that they should be viewed as preliminary.

The nuclear CCSD code presently allows us to incorporate up to eight major oscillator shells within a full calculation. We uncouple all $G$-matrix elements and work with a completely uncoupled (or $M$-scheme in the shell-model parlance) single-particle basis. We checked the code by performing calculations in a fairly small space (the $0s1d$ oscillator space) using both the newly developed CCSD code and standard diagonalization codes. We find exact agreement between the diagonalization and the CCSD for 2 neutrons and the deuteron systems. In all calculations, our two-body amplitudes must obey fermion symmetry such that $t_{ij}^{ab} = -t_{ji}^{ab} = -t_{ij}^{ba} = t_{ji}^{ba}$. In all cases we find these symmetry properties to be exactly preserved. We have also tested that one sign problem or one index problem in any of the one- or two-body correlation amplitude equations will destroy this fermion symmetry property.

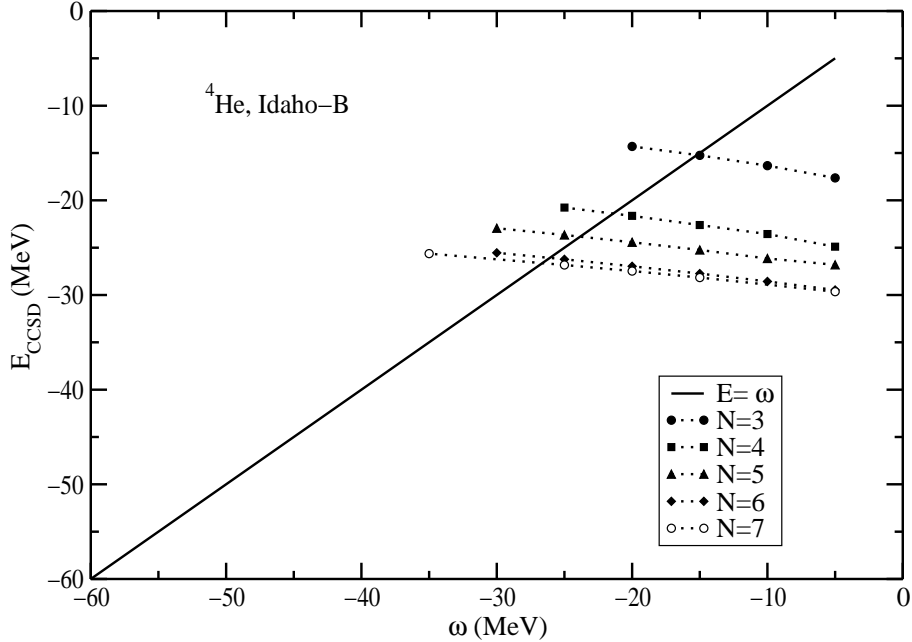

**FIGURE 5.** The CCSD energy for $^4$He as a function of $\omega$ for various model spaces.

We employ the new class of chiral potentials [18] as our bare nucleon-nucleon interaction starting point. The chiral effective Lagrangians employed include one- and two-pion exchange contributions up to chiral order three and contact terms that represent the short-range force. The chiral potential reproduces the $NN$ phase shifts below 300 MeV laboratory energy and the properties of the deuteron with high precision. We use the Idaho-B potential throughout these Proceedings.

Two numerical parameters must be checked with each calculation. One of these is the oscillator parameter $\hbar\omega$. This is a variational parameter in our theory, and we find that the energy is minimal at $\hbar\omega = 11$ MeV for Idaho-B and the $^4$He nucleus. As was mentioned above, the $G$-matrix contains a starting-energy dependence, $\omega$. We know that the energy of the system is given by $E = E(\omega)$. The appropriate choice for $\omega$ is then that energy for which $E = \omega$. We show in Fig. 5 this dependence, along with the dependence of our results on the size of the $P$-space we are considering. Several interesting features emerge from this figure. The first is that as one increases the $P$-space, the resulting energy depends less on $\omega$. This is reasonable: if $P$ were infinite, the solution would recover simply the bare $V$ interaction which has no $\omega$ dependence. The second interesting feature is the rapidity of convergence of the results. Already at seven major oscillator shells one sees the onset of convergence of the total energy. In this model space we obtain the energy $E = -26.6$ MeV. We are currently investigating various possibilities for including the center-of-mass corrections.

## PERSPECTIVES

While the results presented above indicate our first steps toward coupled-cluster theory research, they show outstanding promise. Our $^4$He calculations show evidence of convergence using 7-8 major oscillator shells. Preliminary calculations of $^{16}$O also show convergence within this model space.

Our shopping list for things to pursue in the next 2-3 years is quite long. Since we are just at the beginning of this exciting endeavor, we first want to demonstrate the validity of the method for closed-shell systems such as $^4$He,

$^{16}$O, and $^{40}$Ca. Our immediate challenge is to incorporate a center-of-mass correction for the interaction. The CCSD does not include third-order diagrams, but this deficiency can be alleviated by inclusion of the triples correction [23] transforming our method into CCSD[T]. We will also compare CCSD results which effectively iterates a class of first-, second-, and fourth-order diagrams to all orders, with many-body perturbation theory, which sums all diagrams of a given order. We will explore methods for computing excited states and open-shell systems within CCSD. We will extend CCSD[T] to include three-body interactions. We will also explore the applicability of CCSD[T] to open shell systems and excited-state calculations. We are confident that much can be learned from the many-body physics by moving along this direction of research. We are equally confident that we will eventually be able to extend the coupled-cluster techniques to very neutron-rich nuclei.

## ACKNOWLEDGMENTS

We have had numerous stimulating discussions with several computational chemists and computer scientists including David Bernholdt, Trey White, and Piotr Piecuch. Additionally we acknowledge fruitful conversations with Bruce Barrett, Ray Bishop, Thomas Papenbrock, Michael Strayer, and Niels Walet. This research was sponsored by the Laboratory Directed Research and Development Program of Oak Ridge National Laboratory (ORNL), managed by UT-Battelle, LLC for the U. S. Department of Energy under Contract No. DE-AC05-00OR22725. Computational resources were provided by the National Energy Research Scientific Computing Facility at Laurence Berkeley Laboratory, and the Center for Computational Sciences at ORNL.